\documentclass{article}
\usepackage{spconf,amsmath,graphicx,hyperref}
\usepackage{multirow, multicol}
\usepackage{adjustbox}
\usepackage{enumitem}
\usepackage{amssymb}
\usepackage{booktabs}
\usepackage{bm}
\usepackage{siunitx}
\title{A Unified Multi-Encoder Framework for Music Lyric Intelligibility Assessment: Bridging Invasive and Non-Invasive Paradigms}
\name{Jing Yang\textsuperscript{1,2}, Shenghao Liao\textsuperscript{1}, Shuqing Zhang\textsuperscript{1}, Yongyi Deng\textsuperscript{1}, Pan Li\textsuperscript{1}, Zerui Han\textsuperscript{2}, Gongping Huang\textsuperscript{1*}\thanks{* Corresponding Author.}
\thanks{This work was supported by the National Natural Science Foundation (NSFC) of China under Grant 62471340.}
}
\address{
\textsuperscript{1}School of Electronic Information, Wuhan University, Wuhan Hubei, 430072,  China \\
\textsuperscript{2}MiLM Plus, Xiaomi Inc., Wuhan, China
}
\begin{document}
\ninept
\maketitle
\begin{abstract}
Lyric intelligibility significantly impacts music engagement and comprehension, especially for hearing-impaired individuals. Existing speech intelligibility  metrics perform poorly in musical contexts due to differences in prosody, timing, and spectral dynamics. We propose a unified multi-encoder framework that supports both invasive and non-invasive modeling for robust lyric intelligibility prediction. By capturing complementary local acoustic-semantic alignments, global musical semantics, and ASR-derived textual cues, our method achieves superior performance over baselines on the ICASSP 2026 Cadenza dataset, with the best model achieving 26.61 RMSE, a $9.2$\% improvement over the baseline, demonstrating its flexibility and robustness.

\end{abstract}
\begin{keywords}
Lyric intelligibility, multi-encoder, invasive, non-invasive
\end{keywords}
\section{Introduction}
\label{sec:intro}
Music plays a central role in everyday life, and lyric intelligibility  strongly influences listener comprehension and emotional engagement. For individuals with hearing loss, unclear lyrics often lead to frustration and reduced enjoyment. Traditional speech intelligibility metrics, such as STOI~\cite{5713237} and PESQ~\cite{941023}, are ineffective in musical contexts due to the distinct characteristics of singing, including pitch variation, vocal expression, and stylistic flexibility.

Accurately assessing lyric intelligibility under hearing impairment remains a critical yet underexplored challenge. The ICASSP 2026 Cadenza Challenge~\cite{roa2026cadenza_overview} provides a valuable benchmark with audio clips simulating no, mild, and moderate hearing loss across diverse genres and styles. To address this, we propose a unified multi-encoder framework that supports both invasive (with reference lyrics) and non-invasive (without reference lyrics) modeling. Our method integrates pre-trained models: WavLM~\cite{9814838} for fine-grained acoustic-semantic alignment, MuQ~\cite{zhu2025muqselfsupervisedmusicrepresentation} for global music structure and context, Whisper~\cite{radford2023robust} to generate transcriptions that serve as either soft supervision in the invasive setting or textual input for semantic embedding in the non-invasive setting, and T5~\cite{chung2024scaling} to produce contextual semantic embeddings. By jointly capturing complementary cues through cross-attention and hierarchical fusion, our approach  significantly outperforms baselines on the ICASSP 2026 Cadenza dataset~\cite{roa2025clip}, demonstrating robust generalization.

\section{PROPOSED METHOD}
\label{sec:method}
\begin{figure}[t!]
    \centering
    \includegraphics[width=1.0\linewidth]{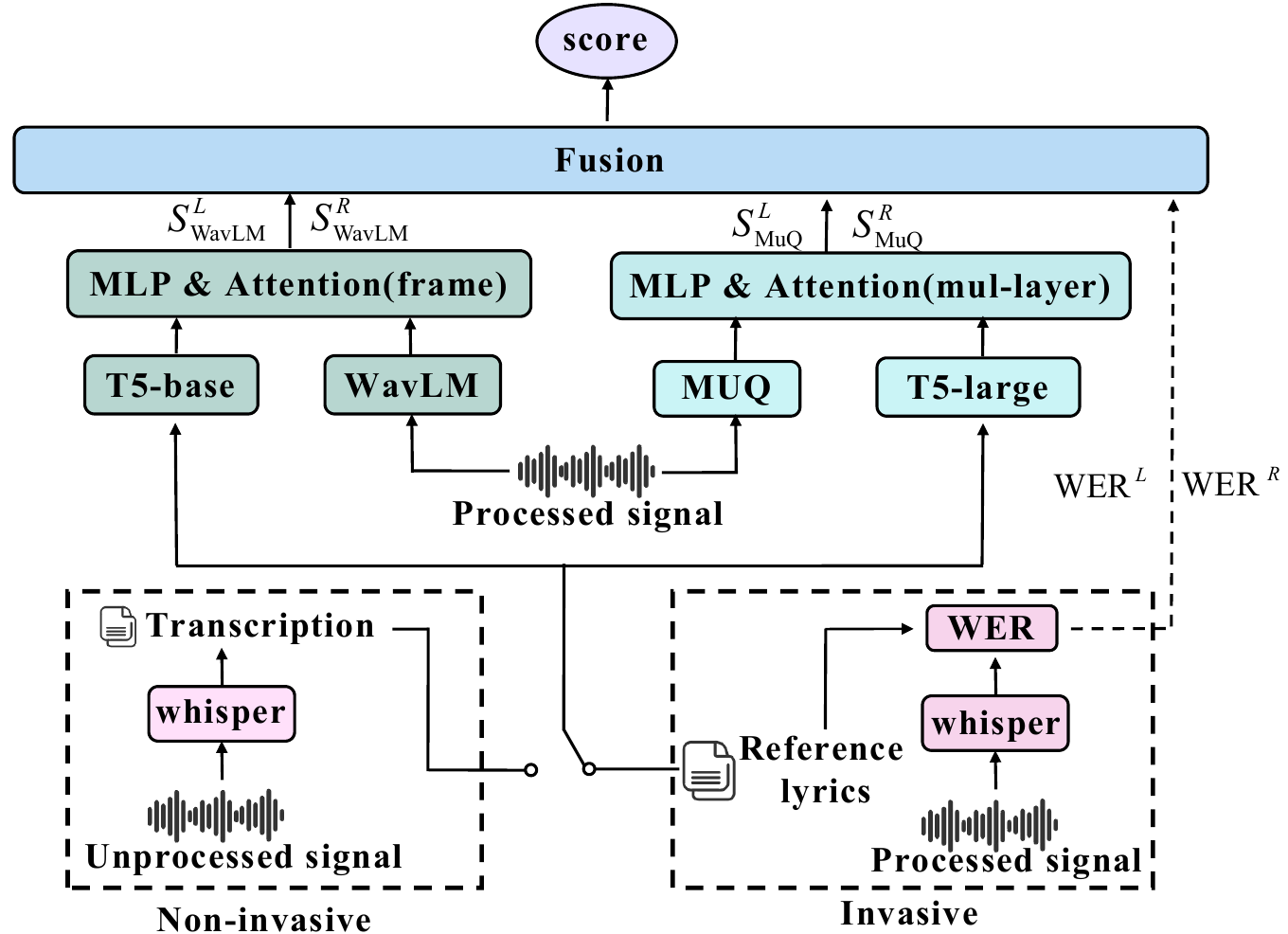}
    \caption{The architecture of the proposed framework, illustrating both invasive (right) and non-invasive (left) modeling paths.}
    \label{fig:invasive}
\end{figure}

We propose a unified multi-encoder framework for lyric intelligibility  assessment, as illustrated in Figure~\ref{fig:invasive}. In the invasive setting, the model takes the processed audio and the corresponding reference lyrics as input. The pre-trained encoders extract complementary acoustic and semantic features through dedicated subnetworks, produce individual score predictions, which are then fused to yield the final intelligibility  assessment. In the non-invasive setting, when reference lyrics are unavailable, the framework substitutes them with ASR-generated transcripts derived from the unprocessed audio and omits the word error rate (WER) computation module, thereby enabling end-to-end training without reference lyrics.

\subsection{Feature Extraction}
% In this work, we extract audio features separately for the left and right ear inputs to account for the asymmetry in the original stereo music. To support dimensionally consistent fusion, we employ different-sized T5 models to generate text embeddings that align with the acoustic features from WavLM  and MuQ.

In this work, we extract audio features separately for the left and right ear inputs to account for the asymmetry in the original stereo music. To ensure dimensional compatibility for fusion, we use T5-base and T5-large to generate text embeddings that match the output dimensions of WavLM and MuQ, respectively.
% We describe each encoder and its role in lyric intelligibility  assessment below.

\textbf{WavLM Encoder.}  
WavLM extracts frame-level acoustic-semantic features from the final layer ($\mathbf{E}_{\text{WavLM}}^L, \mathbf{E}_{\text{WavLM}}^R \in \mathbb{R}^{T_1 \times D_1}$) and aligns them with T5-base text embeddings ($\mathbf{E}_{\text{T5-base}} \in \mathbb{R}^{T_2 \times D_1}$) via cross-attention. This enables fine-grained temporal alignment between singing segments and lyrics, capturing precise acoustic-semantic correspondence as in Equations~\eqref{eq:F_WavLM}--\eqref{eq:S_WavLM_r}.
% \textbf{WavLM Encoder.} 
% We extract the last-layer hidden representations from WavLM as audio features ($\mathbf{E}_{\text{ WavLM}}^R,~\mathbf{E}_{\text{ WavLM}}^L\in \mathbb{R}^{T_1 \times D_1}$), and perform cross-attention with the text embeddings ($\mathbf{E}_{\text{T5-base}} \in \mathbb{R}^{T_2 \times D_1}$) generated by the T5-base model at the frame level. This mechanism dynamically models the temporal alignment between singing segments in the audio and their corresponding lyrical words, effectively capturing the fine-grained acoustic-semantic correspondence, as shown in the following equations:
\begin{align}
 \mathbf{F}_{\text{WavLM}}^R &= \text{Attention}(\mathbf{E}_{\text{T5-base}}, \mathbf{E}_{\text{WavLM}}^R, \mathbf{E}_{\text{WavLM}}^R), \label{eq:F_WavLM} \\
 \mathbf{F}_{\text{T5-base}} &= \text{Attention}(\mathbf{E}_{\text{ WavLM}}^R, \mathbf{E}_{\text{T5-base}}, \mathbf{E}_{\text{T5-base}}), \label{eq:F_T5_base} \\
{S}_{\text{ WavLM}}^R &= \text{MLP}[\mathbf{F}_{\text{WavLM}}^R, \mathbf{F}_{\text{T5-base}}], \label{eq:S_WavLM_r}
\end{align}
where ${S}_{\text{ WavLM}}^R$ denotes the right-ear score, and the same procedure is applied to the left-ear input to obtain the corresponding features and score ${S}_{\text{ WavLM}}^L$.

\textbf{MuQ Encoder.}  
MuQ captures global musical context by fusing low-level acoustic features (layers $1$--$3$) and high-level semantic features (layers $10$--$12$). These representations are concatenated and refined through attention-based pooling to produce global audio embeddings $\mathbf{E}_{\text{MuQ}}^L, \mathbf{E}_{\text{MuQ}}^R \in \mathbb{R}^{L \times D_2}$. 
To align audio and text semantics, we apply cross-attention between them and $\mathbf{E}_{\text{T5-large}} \in \mathbb{R}^{T_3 \times D_2}$, which enhances the model’s robustness in capturing musical context. The per-ear scores $S_{\text{MuQ}}^L$ and $S_{\text{MuQ}}^R$ are computed using the same scoring pipeline as defined in Equations~\eqref{eq:F_WavLM}--\eqref{eq:S_WavLM_r}, with adaptations to accommodate the embedding dimension $D_2$.
% Cross-attention with T5-large embeddings ($\mathbf{E}_{\text{T5-large}} \in \mathbb{R}^{T_3 \times D_2}$) enables semantic alignment between text and audio, enhancing robust modeling of musical context. The per-ear scores $S_{\text{MuQ}}^L$ and $S_{\text{MuQ}}^R$ are derived using the same pipeline as in Equations~\eqref{eq:F_WavLM}--\eqref{eq:S_WavLM_r}, adapted to dimension $D_2$.
% \textbf{MuQ Encoder.} 
% We extract low-level acoustic features (layers $1$–$3$) and high-level semantic features (layers $10$–$12$) from MuQ, concatenate them into a multi-scale representation, and apply attention-based pooling to obtain global audio embeddings %($\mathbf{E}_{\text{MuQ}}^R, \mathbf{E}_{\text{MuQ}}^L \in \mathbb{R}^{L \times D_2}$).
% This representation captures both acoustic structure and high-level musical semantics, which supports lyric intelligibility assessment by capturing global musical context. Cross-attention with T5-large embeddings %($\mathbf{E}_{\text{T5-large}} \in \mathbb{R}^{T_3 \times D_2}$)
% aligns textual context with audio, while the multi-layer design allows the model to capture global cues robust to local distortions, complementing the local alignment capability of WavLM.
% The resulting per-ear scores $S_{\text{MuQ}}^L$ and $S_{\text{MuQ}}^R$ are obtained via the same architectural pipeline as in Equations~\eqref{eq:F_WavLM}--\eqref{eq:S_WavLM_r}, adapted to the MuQ feature dimension. %$D_2$.

\textbf{Whisper Encoder.} 
Whisper, an automatic speech recognition (ASR) system, generates audio transcriptions. In the invasive scenario, where reference lyrics are available, we compute WER between the Whisper output and the true lyrics, using the WER as a soft supervision signal fed into the fusion module. In the non-invasive scenario, without access to reference lyrics, we use the Whisper Large model to generate the transcript, which serves as the textual input. T5 processes this transcript to produce semantic embeddings, which are then fused with WavLM and MuQ features via cross-attention, enabling context-aware alignment without human-annotated lyrics.

\subsection{Fusion Strategy}
\label{fusion strategies}
To integrate the outputs from multiple encoders and modalities, we design distinct fusion strategies tailored to the invasive and non-invasive scenarios, respectively.

\textbf{Invasive Scenario.} We adopt a two-stage fusion approach. Take WavLM as an example, first, we concatenate the left and right ear scores of WavLM with their corresponding WER values, and pass them through a linear layer to obtain per-ear fused scores:
\begin{align} 
S_L &= \text{Linear}([S_{\text{WavLM}}^L, \text{WER}^L]), \\ 
S_R &= \text{Linear}([S_{\text{WavLM}}^R, \text{WER}^R]),
\end{align}
where $S_L$ and $S_R$ represent the fused scores for the left and right ears, respectively. Then the per-ear scores are combined via a weighted sum with learnable weights $\beta_L$ and $\beta_R$ to obtain the final WavLM score $S_{\text{WavLM}}$:
\begin{equation}
S_{\text{WavLM}} = \beta_L \cdot S_L + \beta_R \cdot S_R.
\end{equation}
 Similarly, we compute the final score for MuQ, denoted as $S_{\text{MuQ}}$, using its own per-ear scores and learned weights. Finally, we concatenate $S_{\text{WavLM}}$ and $S_{\text{MuQ}}$ and pass them through a linear layer to produce the final intelligibility  score:
\begin{equation}
S = \text{Linear}([S_{\text{WavLM}}, S_{\text{MuQ}}]).
\end{equation}

\textbf{Non-invasive Scenario.} For each ear, we concatenate the scores from WavLM and MuQ and pass the resulting vector through a linear layer to obtain fused per-ear scores $S_{\text{L}}$ and $S_{\text{R}}$. The final intelligibility  score $S$ is then computed as a weighted sum:
\begin{equation}
S = \alpha_L \cdot S_{\text{L}} + \alpha_R \cdot S_{\text{R}},
\end{equation} 
where $\alpha_L$ and $\alpha_R$ are learnable parameters.

\section{Experimental}
\label{sec:experimental}

\subsection{Experimental Setup}
\label{subsec:setup}
We utilize the pre-trained WavLM~\cite{9814838}, MuQ~\cite{zhu2025muqselfsupervisedmusicrepresentation}, Whisper~\cite{radford2023robust}, and T5~\cite{chung2024scaling} encoders without modifying their core architectures, thereby preserving their pre-trained capabilities. The overall model has approximately $3.3$B total parameters, with only about $6$M trainable parameters in the fusion module, enabling efficient fine-tuning. To compute the WER, we employ multiple Whisper model sizes (small, medium, large) and report the average WER to mitigate variance and improve reliability. The model is trained on a single NVIDIA A800 GPU with a batch size of $4$ for the training set and $2$ for the validation set. Optimization is performed using stochastic gradient descent (SGD) with a momentum of 0.9 and an initial learning rate of \(1 \times 10^{-4}\).

To further validate the design of the multi-encoder fusion strategy, we conduct an ablation study by removing one encoder (WavLM, MuQ, or Whisper) at a time to quantify each component’s contribution. 
% To validate the contribution of each component, we conduct an ablation study by systematically removing one encoder (WavLM, MuQ, or Whisper) at a time,
% %from the full model and evaluating performance on the validation and test sets.
% which quantifies each module’s impact and confirms the effectiveness of the proposed multi-encoder fusion strategy.

\subsection{Results Analysis}
\label{subsec:results}
Table~\ref{tab:results} shows that our method significantly outperforms baselines in both invasive and non-invasive settings. In the validation set, our invasive model reduces RMSE by $9.2$\% and increases correlation by $16.9$\% compared to the Baseline Whisper. In the non-invasive setting, the improvement is even more pronounced: RMSE drops by $14.9$\%, and correlation jumps from $0.21$ to $0.57$ in the evaluation set. The ablation study further confirms the effectiveness of each component: removing Whisper leads to a substantially larger degradation in RMSE and correlation compared to omitting WavLM or MuQ, underscoring Whisper’s dominant role in semantic supervision. Nevertheless, both WavLM and MuQ remain essential for acoustic alignment and global musical context modeling, collectively contributing to the robustness of our multi-encoder framework.

\begin{table}[! t]
    \centering
    \renewcommand{\arraystretch}{0.7} % 调整行间距
    \setlength{\tabcolsep}{3pt}       % 调整列间距
    \caption{Results on the ICASSP 2026 Cadenza dataset}
    \label{tab:results}
    % \begin{tabular}{lllll}
    \begin{tabular}{l S[table-format=2.2] S[table-format=1.2] S[table-format=2.2] S[table-format=1.2]}
    \toprule
     ~ & \multicolumn{2}{c}{validation} & \multicolumn{2}{c}{evaluation}  \\ \midrule
        % System & RMSE$\downarrow$ & Correlation$\uparrow$ & RMSE$\downarrow$ & Correlation$\uparrow$ \\ \midrule
        System & \multicolumn{1}{c}{RMSE$\downarrow$} & \multicolumn{1}{c}{Correlation$\uparrow$} & \multicolumn{1}{c}{RMSE$\downarrow$} & \multicolumn{1}{c}{Correlation$\uparrow$} \\ \midrule
        Baseline Whisper  & 29.32 & 0.59 & 29.08 & 0.58\\ 
        T013a & $\bm{26.61}$ & $\bm{0.69}$ & $\bm{26.54}$ & $\bm{0.67}$\\ 
        T013a-no-MuQ & 27.11  & 0.68 & 27.14  & 0.66\\ 
        T013a-no-WavLM & 28.03  & 0.64 & 27.76  & 0.64\\ 
        T013a-no-Whisper & 30.48  & 0.57 & 30.68  & 0.55\\ \midrule
        Baseline STOI  & 36.11 & 0.14  & 34.89 & 0.21\\
        T013b & 29.75  & 0.59 & 29.68  & 0.57\\ \bottomrule

    \end{tabular}
\end{table}

%\vfill
%\pagebreak

\section{Conclusions}
\label{sec:col}
We proposed a unified multi-encoder framework for lyric intelligibility  assessment, effective in both invasive and non-invasive settings. By leveraging WavLM's frame-level attention for local acoustic-semantic alignment and MuQ's multi-layer pooling for global musical context, our method captures complementary acoustic and semantic cues. Crucially, Whisper's text generation serves as soft supervision via WER in the invasive setting and as textual input for semantic embedding in the non-invasive setting. Evaluated on the ICASSP 2026 Cadenza dataset, our approach achieves superior performance over baselines, validating the effectiveness of the multi-encoder design.

\vfill\pagebreak

\bibliographystyle{ieeetr}
\bibliography{refs}

\end{document}